\RequirePackage{fix-cm}
\documentclass[journal]{IEEEtranTCOM}
\usepackage[latin9]{inputenc}
\usepackage{amsmath}\usepackage{amssymb}\usepackage{graphicx}
\usepackage{tikz}\usetikzlibrary{automata,arrows,positioning,calc}

\makeatletter

\usepackage{epsfig}\usepackage{amsfonts}\usepackage{array}\usepackage{dcolumn}\usepackage{psfrag}\usepackage{amsfonts}\usepackage{accents}\usepackage{booktabs}
\usepackage{cite}\usepackage[all]{xy}
\usepackage{dsfont}\usepackage{url}\usepackage{xcolor}\usepackage[nolist]{acronym}
\usepackage[FIGTOPCAP]{subfigure}
\usepackage[footnote,draft,silent,nomargin]{fixme}
\usepackage{algorithm}
\usepackage[noend]{algpseudocode}
\usepackage{amsthm}
\usepackage{mathtools}
\usepackage{bbm}
\usepackage{xfrac}
\usepackage{booktabs}
\let\labelindent\relax\usepackage{enumitem}
\usepackage{bm}

\acrodef{SINR}{Signal to Interference and Noise Ratio}
\acrodef{CDF}{Cumulative Distribution Function}
\acrodef{PDF}{Probability Density Function}
\acrodef{MTC}{Machine-type Communications}
\acrodef{M2M}{Machine-to-Machine}
\acrodef{CTMC}{Continuous-Time Markov Chain}
\acrodef{RV}{Random Variable}
\acrodef{MTTR}{Mean Time to Restoration}
\acrodef{TTI}{Transmission Time Interval}
\acrodef{SDN}{Software Defined Networking}

\renewcommand{\vec}[1]{\underline{#1}}

\newcommand{\figscl}{0.9}
\makeatother

\begin{document}

\title{Ultra-Reliable Low Latency Communication (URLLC) using Interface Diversity}

\author{Jimmy~J.~Nielsen,~\IEEEmembership{Member,~IEEE,}
		Rongkuan~Liu,~
		and Petar~Popovski,~\IEEEmembership{Fellow,~IEEE}
\thanks{Jimmy J. Nielsen and Petar Popovski are with the Department of Electronic Systems, Aalborg University, 9220 Aalborg, Denmark, e-mail: jjn@es.aau.dk, petarp@es.aau.dk}
\thanks{Rongkuan Liu is with the Communication Research Center, Harbin Institute of Technology, Harbin 150001, China, e-mail: \mbox{liurongkuan@hit.edu.cn}.}
}

\maketitle
\begin{abstract} 
An important ingredient of the future 5G systems will be Ultra-Reliable Low-Latency Communication (URLLC). A way to offer URLLC without intervention in the baseband/PHY layer design is to use \emph{interface diversity} and integrate multiple communication interfaces, each interface based on a different technology. In this work, we propose to use coding to seamlessly distribute coded payload and redundancy data across multiple available communication interfaces. We formulate an optimization problem to find the payload allocation weights that maximize the reliability at specific target latency values. 
In order to estimate the performance in terms of latency and reliability of such an integrated communication system, we propose an analysis framework that combines traditional reliability models with technology-specific latency probability distributions.
Our model is capable to account for failure correlation among interfaces/technologies. By considering different scenarios, we find that optimized strategies can in some cases significantly outperform strategies based on $k$-out-of-$n$ erasure codes, where the latter do not account for the characteristics of the different interfaces. The model has been validated through simulation and is supported by experimental results. 
\end{abstract}
\begin{IEEEkeywords}
Communication system reliability, diversity methods, redundancy, codes, real-time systems.
\end{IEEEkeywords}

\section{Introduction}
The upcoming 5G technology is designed for three main use cases, namely enhanced Mobile Broadband (eMBB), massive Machine-Type Communications (mMTC), and Ultra-Reliable and Low Latency Communication (URLLC) \cite{carvalho2016random}. URLLC may be supported both through the 5G new air interface \cite{ji2017introduction} or through the integration of different existing communication technologies \cite{andrews2014will} \cite{monserrat2015metis}.  URLLC will enable the support of new use cases under the umbrella of mission critical \ac{MTC}, whose requirements exceed the capabilities of current wireless technologies. Reliability requirements in terms of packet delivery success rates may be as high as 5-nines ($1\!-\!10^{-5}$) to 9-nines ($1\!-\!10^{-9}$), while also the acceptable latency may be at the sub-second level or even down to a few milliseconds \cite{ratasuk2015recent}. There are proposals for how to decrease the latency in future cellular systems, e.g., by reducing the \ac{TTI} \cite{lahetkangas2014achieving,tullberg2014towards}, fast uplink access \cite{3GPPTR-36881}, or by puncturing URLLC resources on top of eMBB \cite{ji2017introduction}. However, the benefits of such improvements cannot be reaped until the features have been widely rolled out. Furthermore, very high levels of reliability are difficult to achieve with any single wireless communication technology, and is as such expected to be reachable through the integration of multiple communication technologies \cite{dahlman20145g}. 

The use of multiple communication technologies is conceptually very similar to many existing multipath protocols that increase end-to-end reliability~\cite{qadir2015exploiting}. However, the strict latency requirements of mission critical MTC, exclude protocols that rely on retransmission.  Instead we focus on \emph{interface diversity} which is in fact path diversity \cite{apostolopoulos2000reliable}, where each path must use a different communication interface. While there are many multipath protocols \cite{qadir2015exploiting}, we have not identified any works that allow to flexibly trade-off latency and reliability, as considered in this paper. The closest examples of related work that we have identified are the following. In \cite{yap2012making}, the authors demonstrate the use of \ac{SDN} to distribute application packets across multiple available interfaces to increase application throughput. This work is extended in \cite{yap2013scheduling} by proposing a load balancer that also takes the user's preferences into account when selecting interfaces for different applications' packets. In \cite{singh2016optimal}, the authors present an analysis of multi-link aggregation in heterogeneous wireless systems. Specifically, they optimize the network utility (and throughput) for a specified degree of multi-user fairness. Candidate architectures for enabling multi-connectivity and high reliability in 3GPP cellular systems are studied in \cite{michalopoulos2016user} and \cite{ravanshid2016multi}. Most recently, in \cite{wolf2017diversity}, the authors present a physical layer analysis of outage probability in multi-connectivity scenarios.

In this work we are focusing both on achieving ultra high reliability by using multiple interfaces simultaneously and on exploring the potential for reducing latency by splitting the total amount of information to transmit across different interfaces. We demonstrated these principles and the analysis framework in previous work \cite{nielsen2016latency} and explored them in more details in later work \cite{nielsen2017latency}. The present manuscript is a coherent and expanded presentation of the concept of interface diversity for URLLC.

In this paper, we present our proposed analysis framework for estimating the latency and reliability performance of different interface diversity strategies.
The framework uses traditional reliability engineering methods for calculating the reliability of a multi-interface system, given interface specific latency-reliability characteristics.
 Furthermore, we demonstrate how coding can be exploited to enable flexible splitting of payload across interfaces in order to trade-off reliability, packet transmission latency, and bandwidth usage. Increasing the amount of coded information being transmitted on different interfaces between the source device and remote host, generally increases the probability of successful reception. However, the increased payload size also incurs an increase in latency, i.e., the time from a message is generated in the source device, until it is successfully received in the remote host. Also, transmitting more information results in a larger bandwidth consumption. For studying this trade-off, we formulate the optimization problem of the optimal payload splitting problem as well as the generic evaluation method and present corresponding numerical results. For the specific case of splitting data between two interfaces we provide an analytic solution to minimize the expected latency.
For evaluating the performance of systems with correlated interface failures, we propose a Markov model that jointly accounts for the technology-specific latency-reliability characteristics and infrastructure failure/restoration probabilities and dependencies. While the proposed Markov chain is specific to the considered use case, the presented modeling principle can be applied to other system configurations. This Markov model is a significant revision of the model used in~\cite{nielsen2016latency}.

The paper is organized as follows. 
In sec.~\ref{sec:system_model} we introduce the MTC system model and interface diversity transmission strategies. 
Analysis and modeling of the transmission strategies is presented in sec.~\ref{sec:reliability_miftx} and \ref{sec:miftx_correlated} for the cases of uncorrelated and correlated failure models, respectively. We present and discuss the numerical results in sec.~\ref{sec:results}. Finally, conclusion and outlook are given in sec.~\ref{sec:conclusion}.


\begin{figure}
    \centering
    \subfigure[Latency-reliability function]{\includegraphics[width=0.47\textwidth]{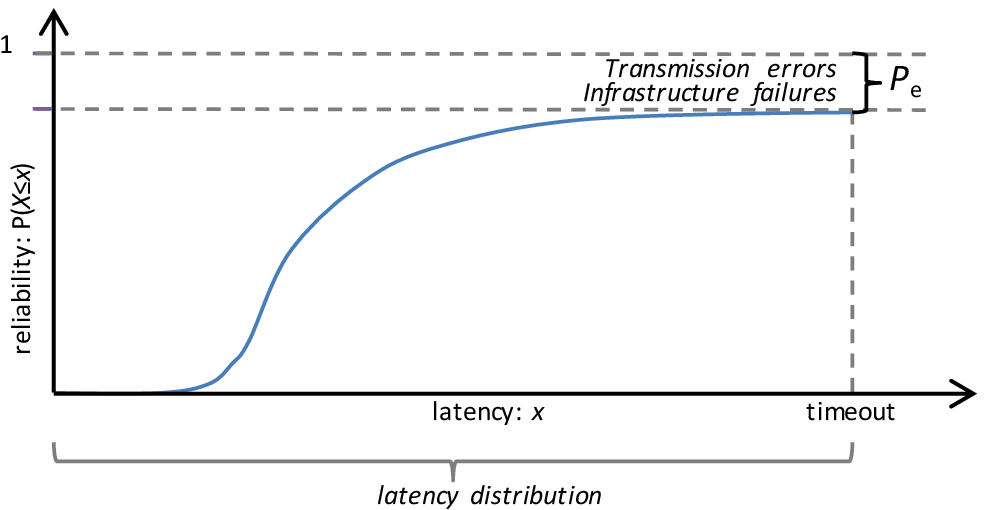}}
    ~ 
    \subfigure[Network diagram]{\includegraphics[width=0.47\textwidth]{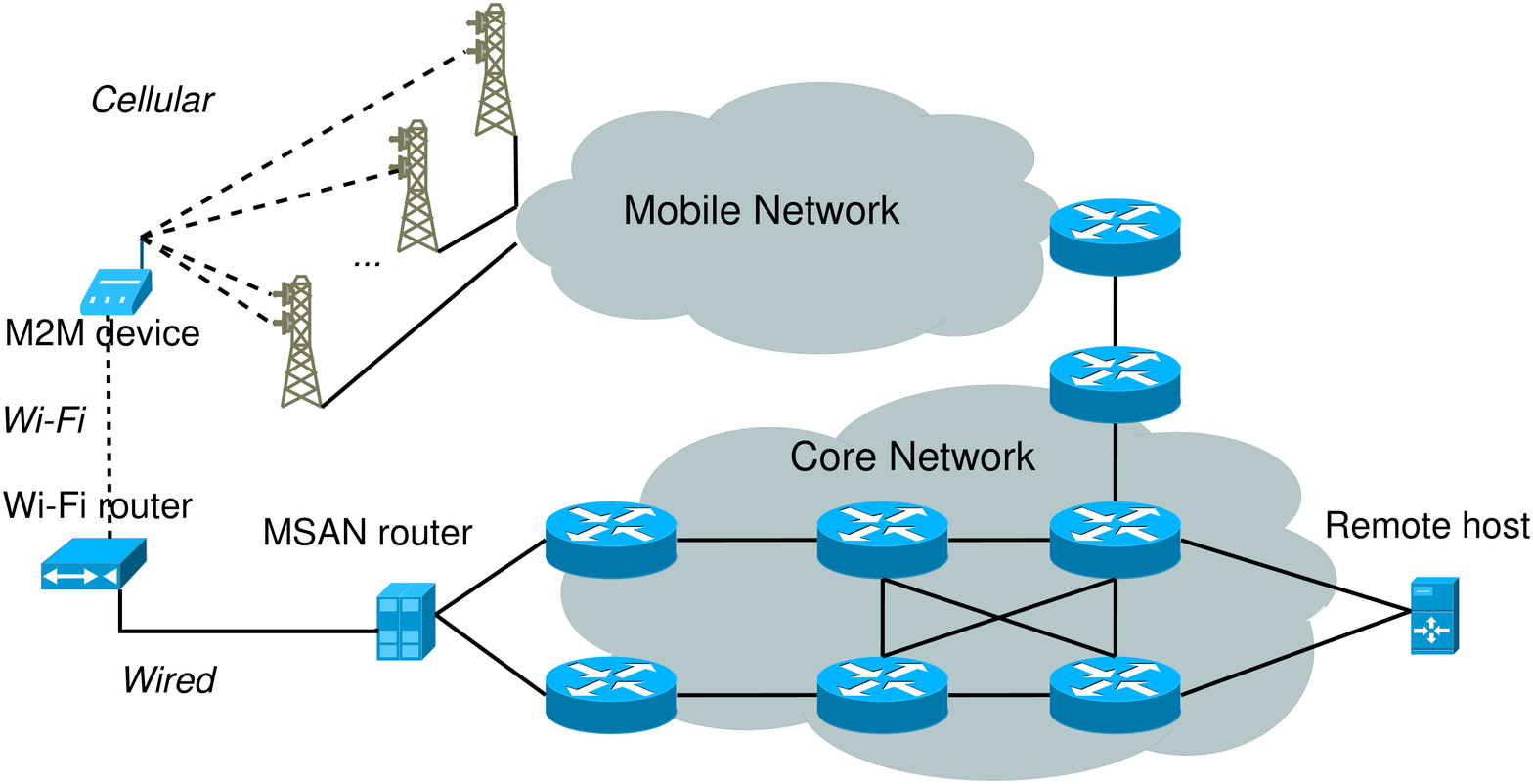}}
    \caption{(a) Conceptual latency-reliability function. (b) Multiple paths between M2M device (left) and remote host (right).}
    \label{fig:system_model}
\end{figure}

\section{System model}
\label{sec:system_model}
We consider an M2M device that needs to communicate reliably with a specific end-host, e.g., a monitoring device reporting measurements, status and alarm messages to a control unit.
The M2M device has $N$ communication interfaces (wired and cellular) available to reach the end-host. An example of a deployment with two cellular and one wired interface is depicted in Fig. \ref{fig:system_model}(b). Notice that some interfaces that are physically separated are subject to (almost) independent failures, while cellular connections that share the same base station may have a higher degree of failure correlation. When transmitting information through the different communication interfaces, the individual messages will be subject to varying delays and packet losses, which can be characterized with the latency-reliability function \cite{strom20155g} exemplified in Fig. \ref{fig:system_model}(a). For our analysis we assume that the latency-reliability functions of the interfaces are available, being previously obtained from network monitoring measurements of end-to-end delay.



\subsection{Transmission Strategies}\label{sec:strategies}
\begin{figure}[h]
    \centering
    \subfigure[Cloning]{\includegraphics[width=0.37\textwidth]{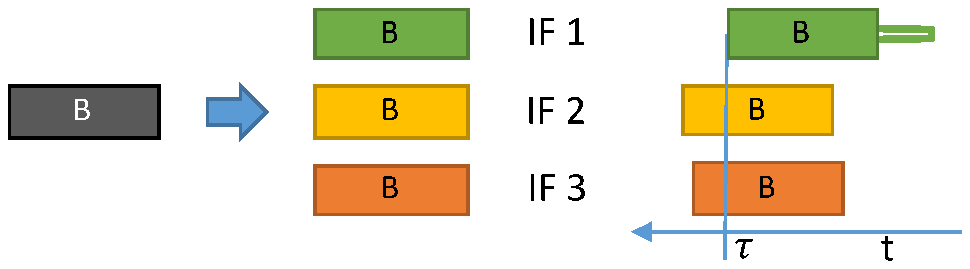}}
    ~ 
    \subfigure[2-out-of-3]{\includegraphics[width=0.37\textwidth]{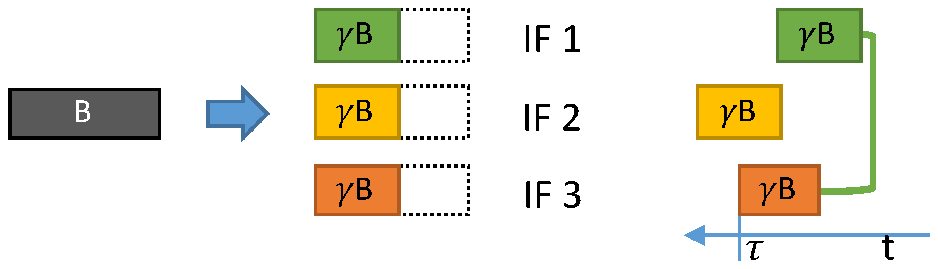}}
    \subfigure[Weighted]{\includegraphics[width=0.37\textwidth]{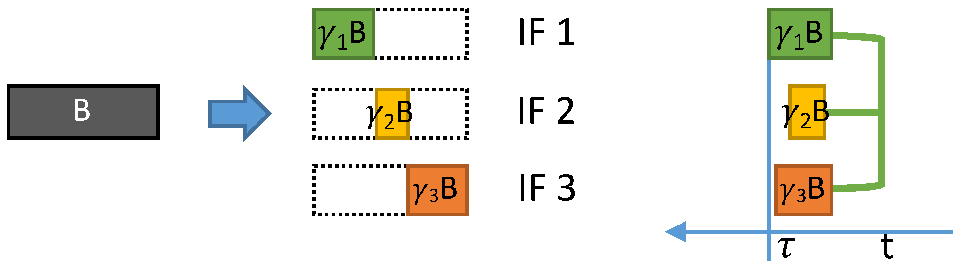}}
    \caption{Transmission strategies, with 2-out-of-3 as example of $k$-out-of-$N$. The time instant $\tau$ is when the payload can be successfully decoded.}
    \label{fig:strategies}
\end{figure}

We consider the following three strategies, for transmitting the stream of messages from the M2M device to the end-host (see Fig. \ref{fig:strategies}):
\subsubsection{Cloning} In this simple approach, the source device sends a full copy of each message through each of the $N$ available interfaces. Since only one copy is needed at the receiver to decode the message, cloning makes the communication robust at the expense of $N-$fold redundancy.
\subsubsection{Splitting} Covers the types of strategies where instead of sending a full copy on each interface, only a fraction of the message is sent on each interface. This allows to trade-off reliability and latency through the selection of the fraction sizes. 
While a gain in reliability can always be achieved by sending more redundancy information, a reduction of latency is not always possible to achieve due to the following. The end-to-end delay of a data transmission is in the considered type of scenarios, primarily determined by the wireless access protocol, and consists of a protocol-dependent access latency, $t_\text{a}$, and the actual time it takes to transfer the (coded) payload, $t_\text{t}$, which is a function of the bitrate. Simply put: $t_\text{e2e} = t_\text{a} + t_\text{t}$. When using a splitting strategy, we are only able to reduce $t_\text{t}$. For small packets $t_\text{a} >> t_\text{t}$, there is no noticeable gain if we reduce $t_\text{t}$. However, for large packets, when $t_\text{a} << t_\text{t}$, splitting can help to reduce latency.

We assume that the payload is encoded, such that we can generate a desired number of coded fragments to be sent through different interfaces. This can be achieved using for example rateless codes \cite{mackay2005fountain} or Reed Solomon codes \cite{wicker1999reed}.
The receiver will be able to decode the encoded message with very high probability as long as it receives coded fragments corresponding to approximately $100 (1+\epsilon) \%$ of the initial message size. A typical value is $\epsilon=0.05$ \cite{mackay2005fountain} and we denote this threshold as $\gamma_\text{d}=1.05$. The coded fragments of a message that are to be sent over the same interface, are grouped together in a single packet to avoid excess protocol overhead.
We assume that for a specific payload message, we let the used code (e.g. rateless or Reed Solomon based) generate coded fragments of a relatively small size, e.g. 10 bytes. When nonuniform, \emph{weighted} splitting is used, the challenge is to determine how many fragments to assign to each interface.
Depending on whether identical or different types of interfaces are used, splitting can be realized through either $\bm{k}$-out-of-$\bm{N}$ splitting or weighted splitting, respectively:
	\begin{description}
		 \item [{$\bm{k}$-out-of-$\bm{N}$}] splitting generates $n$ equally sized coded fragments from the payload and the receiver needs to receive at least $k$ of them in order to  decode the message. This strategy allows to trade off reliability and latency, since large redundancy leads to higher reliability but longer transmission times, whereas small redundancy offers a lower error protection but shorter transmission times.
		\item [{Weighted}] the payload is split across interfaces so that the size of the per-interface packet is optimized according to a specific objective. That objective could be to minimize the expected overall transmission latency or to maximize the reliability for a given latency constraint. The optimal solution is, however not trivial, as our analysis shows.
	\end{description}

\subsection{Achievable Latency Reduction}
When using splitting, the  $t_t$ for the different interfaces determine the optimal $\gamma$ and thereby how much the latency can be reduced. In principle, for infinitely large payloads and identical interfaces, the latency can be reduced to $1/N$ of a single interface's latency. In practice, payload sizes are limited and interfaces may have different characteristics. In the following we analyze the achievable latency reduction for the simple case of a two interface system.
Let $t_t^{(1)}$ and $t_t^{(2)}$ be the transmission latencies of two interfaces. Using cloning the E2E latency is $t_\text{E2E}^\text{clon} = \min(t_\text{t}^{(1)}+t_\text{a}^{(1)},t_\text{t}^{(2)}+t_\text{a}^{(2)})$ and when splitting the coded payload between the two interfaces, the latency is $t_\text{E2E}^\text{split} = \max(\tilde{t_{\text{t}}}^{(1)}+t_\text{a}^{(1)},\tilde{t_{\text{t}}}^{(1)}+t_\text{a}^{(2)})$, where $\tilde{t_{\text{t}}}^{(i)}$ are the transmission latencies when splitting the coded payload.
Consequently, the latency reduction is: $G_\text{E2E} = \frac{t_\text{E2E}^\text{clon}-t_\text{E2E}^\text{split}}{t_\text{E2E}^\text{clon}}$.

Let $\gamma$ be the fraction of coded data sent via interface 1 and $1-\gamma$ be the fraction of coded data sent via interface 2. 
The optimal choice of the fraction $\gamma$ is for the non-stochastic case calculated as: $\gamma = (t_\text{t}^{(2)}+t_\text{a}^{(2)})/(t_\text{t}^{(1)}+t_\text{a}^{(1)}+t_\text{t}^{(2)}+t_\text{a}^{(2)})$.
Then, the transmission times of the two interfaces with splitting are $\tilde{t_{\text{t}}}^{(1)} = \gamma_\text{d} \gamma t_\text{t}^{(1)}$ and $\tilde{t_{\text{t}}}^{(2)} = \gamma_\text{d} (1-\gamma) t_\text{t}^{(2)}$.

\begin{figure}[h]
    \centering
    \subfigure[$t_\text{a}^{(1)}=t_\text{a}^{(2)}=1$]{\includegraphics[width=0.45\textwidth]{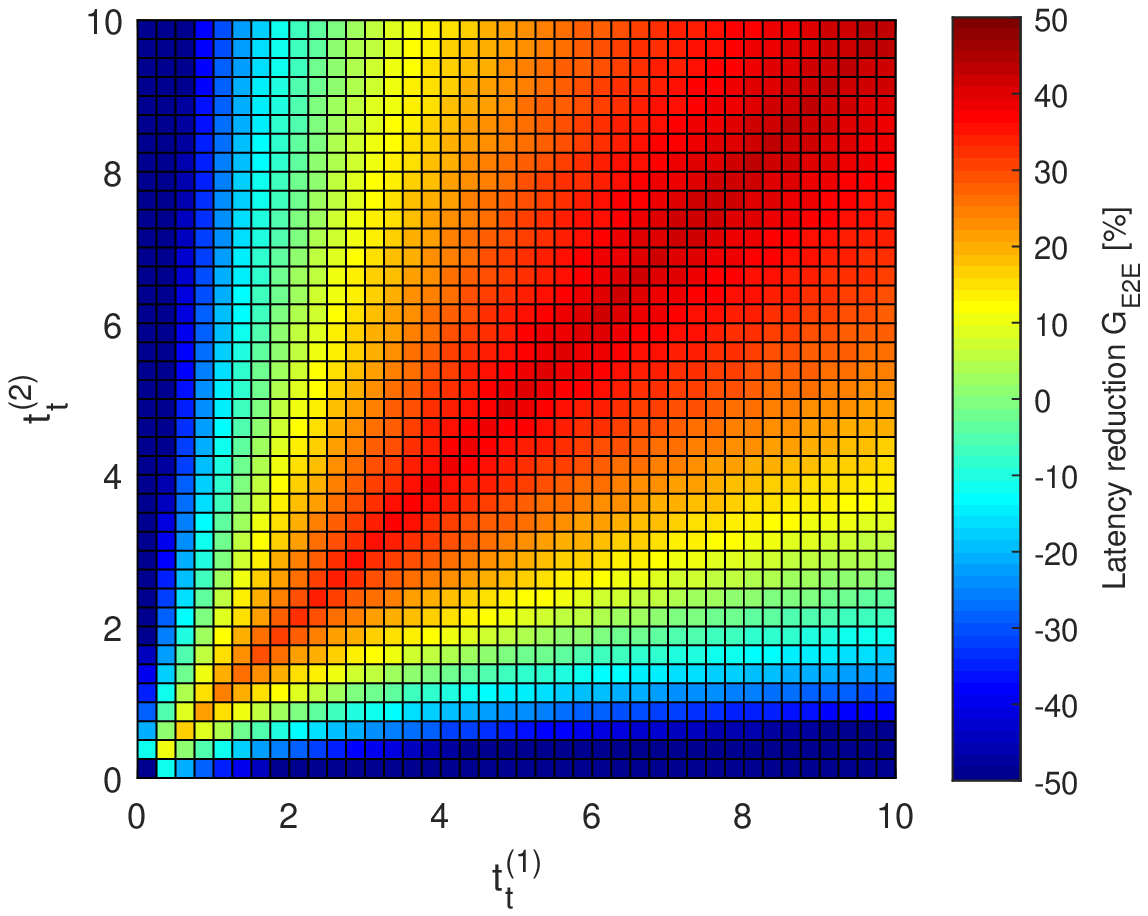}}
    ~ 
    \subfigure[$t_\text{a}^{(1)}=0.5$, $t_\text{a}^{(2)}=1$]{\includegraphics[width=0.45\textwidth]{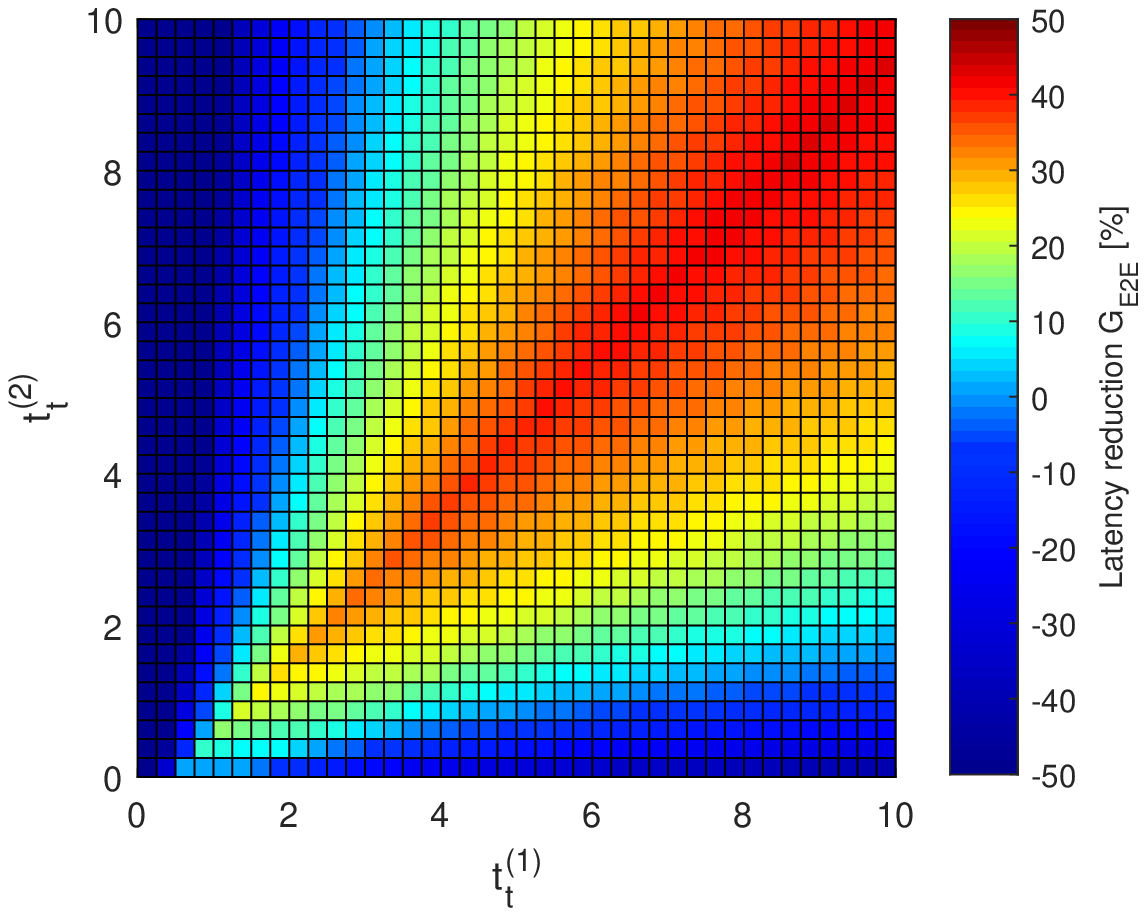}}
    \caption{Achievable latency reduction $G_\text{E2E}$ of two interface splitting for different combinations of transmission latency $t_\text{t}^{(1)}$ and $t_\text{t}^{(2)}$.}
    \label{fig:latency_tradeoff}
\end{figure}

In Fig. \ref{fig:latency_tradeoff} we have plotted the achievable latency reduction of different combinations of transmission latencies, for two cases, namely when the two interfaces have the same access latency $t_\text{a}$ and the case where the access latency of interface 2 is twice that of interface 1. When the access latencies are the same, a close to 50\% latency reduction is possible when the transmission latencies $t_\text{t}^{(1)}$ and  $t_\text{t}^{(2)}$ are also equal and 6-10 times larger than the access time $t_\text{a}$. Using the numbers presented later in Table \ref{tab:pl_to_rss_and_reliability}, we find that such ratios occur for GPRS, EDGE, UMTS, and HSDPA with large payloads of 3000 bytes or more. In Fig. \ref{fig:latency_tradeoff}(b) we see that when access latencies are different, the range of transmission latencies that lead to latency reductions is more narrow. While this could seem like a worse result, one should keep in mind that since the starting point was a lower latency on interface 1, the resulting E2E latency in case (b) is still lower than in case (a), even though the relative latency reduction is less.

In conclusion, one should keep in mind that while splitting reduces latency, it simultaneously sacrifices reliability. In the following, we present an analytic framework that uses the latency probability distribution to quantify the interplay between latency and reliability, which we in turn use to study different deployment scenarios.

\subsection{Latency-reliability Function}
As the duration of a packet transmission is usually depending on the packet size, it is necessary to characterize the relationship between the payload size and the latency distribution. 
Let $F_i(x,B)$ denote the latency-reliability function of the $i-$th interface, which is  the probability of being able to transmit a data packet of $B$ bytes from a source to a destination via interface $i$ within a latency deadline of $x$. 
In other words, the value of $F_i(x,B)$ is the achievable reliability ($P(X \leq x)$) for a latency value $x$ and payload size $B$.
In the following, we let $\gamma_i$ specify the fraction of payload assigned to interface $i$, where $\gamma_i=[0,\gamma_\text{d}]$. 
The notation $P_\text{e}^{(i)}$ refers to $P_\text{e}$ (defined in Fig. \ref{fig:system_model}(a)) for the $i-$th interface.

In this work, we assume that the latency-reliability functions are static for each considered interface, meaning that the applied transmissions strategies are not dynamically changed. In reality, there will be variations and error bursts over time. But without a reliable means for predicting such fluctuations before they occur, it will be impossible to achieve ultra-reliable operation, since just a few errors or spikes in latency can be catastrophic in the ultra-reliable domain. We therefore leave the dynamic policy selection as a future work item.

\section{Reliability of transmissions over independent interfaces}\label{sec:reliability_miftx}
This section presents the proposed methodologies for achieving reliability through diversity of independent interfaces, i.e. interfaces that do not have common error causes.

\subsection{Evaluating reliability for weight assignment}\label{sec:evaluating_weight}
The general approach to evaluating the latency-reliability function for a specific transmission strategy can be described as follows. The success probability is calculated by summing up the probability of \emph{successful outcomes}. A successful outcome is a combination of lost and received coded packets, for which the receiver can successfully decode the original message. This is further explained below.

Evidently, the payload assignments with $\sum_{i=1}^N \gamma_i < \gamma_\text{d}$ should be avoided, as they can unlikely lead to a successful decoding outcome. 
For enumeration of all possible outcomes we use the $2^N \times N$ matrix $\mathbf{C}$:
\begin{equation}
	\textbf{C} = \begin{bmatrix}
		0 		& 0 	 & \cdots & 1 \\
		\vdots 	& \vdots & \cdots & \vdots \\
		0 		& 1 	 & \cdots & 1 \\
	\end{bmatrix}^\text{T}.
\end{equation}
The element $c_{h,i}$ in the $h$th row and $i$th column of $\mathbf{C}$ is 0/1 if the $h-$th possible outcome features a successful/failed reception over the $i-$th interface.

For a specific $\bm{\gamma}$, we use the law of total probability to evaluate the resulting latency-reliability:
\begin{equation}\label{eq:weighted_eval}
	F_\text{weighted}(x,\bm{\gamma},B) = \sum\limits_{h=1}^{2^N} d_h \prod\limits_{i=1}^N G_i(x,\gamma_i B)
\end{equation}
where 
\begin{equation}
		d_h = \left\{  \begin{array}{lr}
		1, & \text{if } \sum_{i=1}^N c_{h,i} \cdot \gamma_i \geq \gamma_d\\
		0, & \text{otherwise}
		\end{array}\right.
\end{equation}
ensures that we only include successful outcomes. Furthermore, $G_i(x)$ is defined as:
\begin{equation}
		G_i(x,\gamma_i B) = \left\{  \begin{array}{ll}
		F_i(x,\gamma_i B), & \text{if } c_{h,i} = 1\\
		1-F_i(x,\gamma_i B), & \text{if } c_{h,i} = 0 .
		\end{array}\right.
\end{equation}
We note that the product $\prod_{i=1}^N G_i(x,\gamma_i B)$ in eq. \eqref{eq:weighted_eval} occurs as a CDF of a maximal value of $N$ random variables, since the latency of the decoding corresponds to the last arriving segment (maximal time) that enables successful decoding.

\subsection{Cloning}
For transmissions using packet cloning over $N$ interfaces that can justifiably be considered independent, e.g., Wi-Fi and cellular or cellular from different operators, we can either use the method presented above or we can use the easier traditional parallel systems \cite{rausand2004system} method to combine the latency-reliability functions as:
\begin{equation}
	F_\text{$N$-clon}(x,\bm{\gamma}, B) = 1-\prod\limits_{i=1}^N (1-F_i(x,\gamma_i B))\label{eq:f_k_par}.
\end{equation}
In either case $\gamma_i=1$ for $i={1, \ldots , N}$.

\subsection{$k$-out-of-$N$ splitting}
While the $k$-out-of-$N$ splitting strategy is only optimal for the case of identical interfaces, it can in principle be used in any case, but with best results in situations where the properties of the available interfaces are comparable.
We can evaluate the latency-reliability function using eq. \eqref{eq:weighted_eval} with $\gamma_i = \sfrac{1}{k}$ for $i={1, \ldots, N}$.

\subsection{Weighted splitting between two interfaces}\label{sec:analysis}
Initially, we analyze the simplest case of weighted splitting, where we have only two interfaces. Specifically, we consider how to optimally split coded payload between two interfaces A and B, so that latency is minimized. 
For this, we formulate an analytical solution to a subproblem of the general weighted splitting optimization problem that is presented in the subsequent subsection.

In the two-interface optimization problem, we assume the latency of each interface is represented by two Gaussian random variables $ X_{A} \sim \mathcal{N} (\mu_{A}, \sigma_{A}^{2})$ and $ X_{B} \sim \mathcal{N} (\mu_{B}, \sigma_{B}^{2})$. In the following we assume that $\sigma_{A}$ and $\sigma_{B}$ are constant and independent of $\mu_{A}$ and  $\mu_{B}$.
When splitting the payload between two interfaces, latency is defined as the time at which the last fragment is received. The expected latency is thus the expectation of $\max (X_{A},X_{B})$, which is also the first moment of the random variable $\max (X_{A},X_{B})$. 

By using analytical approximation of the expectation of the maximum of two normal random variables \cite{clark1961greatest}, we obtain:
\begin{align}
L = \mathbb{E}[ \max (X_{A},X_{B}) ] = \mu_{A} \Phi (\eta) + \mu_{B} \Phi (-\eta) + \xi \phi (\eta)
\end{align}
where $\phi(x)\!=\!\frac{1}{\sqrt{2 \pi}} \exp^{ -\frac{x^2}{2} }$, $\Phi (x)\!=\!\int_{-\infty}^{x} \phi (t) \mathrm{d}t $,
$\eta\!=\!\frac{ \mu_{A}-\mu_{B} }{ \xi }$, and $ \xi\!=\!\sqrt{ \sigma_{A}^{2} + \sigma_{B}^{2} } $.

To find the minimum of the expected latency, we differentiate $L$ with respect to $\gamma$:
\scalebox{0.825}{\parbox{.5\linewidth}{%
\begin{align}
\frac{\mathrm{d}L}{\mathrm{d}\gamma} &= \frac{\mathrm{d}\mu_{A}}{\mathrm{d}\gamma} \Phi (\eta) + \mu_{A} \phi (\eta) \frac{\mathrm{d}\eta}{\mathrm{d}\gamma} + \frac{\mathrm{d}\mu_{B}}{\mathrm{d}\gamma} \Phi (-\eta) - \mu_{B} \phi (-\eta) \frac{\mathrm{d}\eta}{\mathrm{d}\gamma} + \xi \phi^{\prime} (\eta) \frac{\mathrm{d}\eta}{\mathrm{d}\gamma} \notag \\
&= \frac{\mathrm{d}\mu_{A}}{\mathrm{d}\gamma} \Phi (\eta) + \frac{\mathrm{d}\mu_{B}}{\mathrm{d}\gamma} \Phi (-\eta) + (\mu_{A} \phi (\eta) - \mu_{B} \phi (-\eta) + \xi \phi^{\prime} (\eta)) \frac{\mathrm{d}\eta}{\mathrm{d}\gamma}\notag.
\end{align}
}}
Since $\mu_{A} \phi (\eta) - \mu_{B} \phi (-\eta) + \xi \phi^{\prime} (\eta) = 0$, and by using the definition of $\mu$ from eq. \eqref{eq:latency_B} we obtain:
\scalebox{0.95}{\parbox{.5\linewidth}{%
\begin{equation}
\frac{\mathrm{d}L}{\mathrm{d}\gamma} = \frac{\mathrm{d}\mu_{A}}{\mathrm{d}\gamma} \Phi (\eta) + \frac{\mathrm{d}\mu_{B}}{\mathrm{d}\gamma} \Phi (-\eta) = \frac{\alpha_{A}}{2} \Phi (\eta) - \frac{\alpha_{B}}{2} \Phi (-\eta).
\end{equation}
}}

In order to get the optimal solution, $\frac{\mathrm{d}L}{\mathrm{d}\gamma} = 0$ must hold.
So we have the solution as follows:
\begin{align}
\left\{
\begin{array}{lcc}
\Phi (-\eta) = \frac{\alpha_{A}}{\alpha_{A}+\alpha_{B}}, & \mbox{if} & \eta \geq 0 \notag \\
\Phi (\eta) = \frac{\alpha_{B}}{\alpha_{A}+\alpha_{B}}, & \mbox{if} & \eta < 0 \notag
\end{array}
\right.
\end{align}
which is equivalent to:
\begin{align}\label{eq:analytic_splitting}
\left\{
\begin{array}{lcc}
\gamma= \frac{\alpha_{B} + \beta_{B} - \beta_{A} - 2 \xi \Phi^{-1} (\frac{\alpha_{A}}{\alpha_{A}+\alpha_{B}}) }{ \alpha_{A} + \alpha_{B} }, & \mbox{if} & \mu_{A} \geq \mu_{B} \\
\gamma= \frac{\alpha_{B} + \beta_{B} - \beta_{A} + 2 \xi \Phi^{-1} (\frac{\alpha_{B}}{\alpha_{A}+\alpha_{B}}) }{ \alpha_{A} + \alpha_{B} }, & \mbox{if} & \mu_{A} < \mu_{B}.
\end{array}
\right.
\end{align}

\subsection{Weighted splitting}
Generally, the challenge of the weighted splitting scheme is to determine how many coded fragments to send on each interface to optimize a given utility function. This problem has $N$ degrees of freedom in the form of the payload allocation vector $\bm{\gamma}=\{\gamma_1, \ldots, \gamma_N\}$. 
Formally, this optimization problem can be phrased in the following way:
\begin{equation}\label{eq:opt_weighted}
\begin{array}{rl}
	\underset{\bm{\gamma}}{\arg\max}  & \sum\limits_{r=1}^R F_\text{weighted}(l_r,\bm{\gamma}) \cdot w_r \\
	\text{s.t.} 		& \gamma_i \leq \gamma_\text{d} \\
						& \sum\limits_{i=1}^N \gamma_i \geq \gamma_\text{d}.\\
\end{array}
\end{equation}
where $F_\text{weighted}(l_r,\bm{\gamma})$ is evaluated using eq. \eqref{eq:weighted_eval} and the vectors $\mathbf{l}=\{ l_1, \ldots, l_R\}$ and $\mathbf{w}=\{ w_1, \ldots, w_R\}$ specify the targeted latency values to be maximized and their corresponding importance, respectively. For example, $\mathbf{l}=\{ 0.2, 0.5\}$ and $\mathbf{w}=\{ 1, 10\}$ would mean that reliability at 0.5 s is 10x more important than reliability at 0.2 s.

\begin{figure}
    \centering
    \subfigure[Interface curves]{\includegraphics[width=0.4\textwidth]{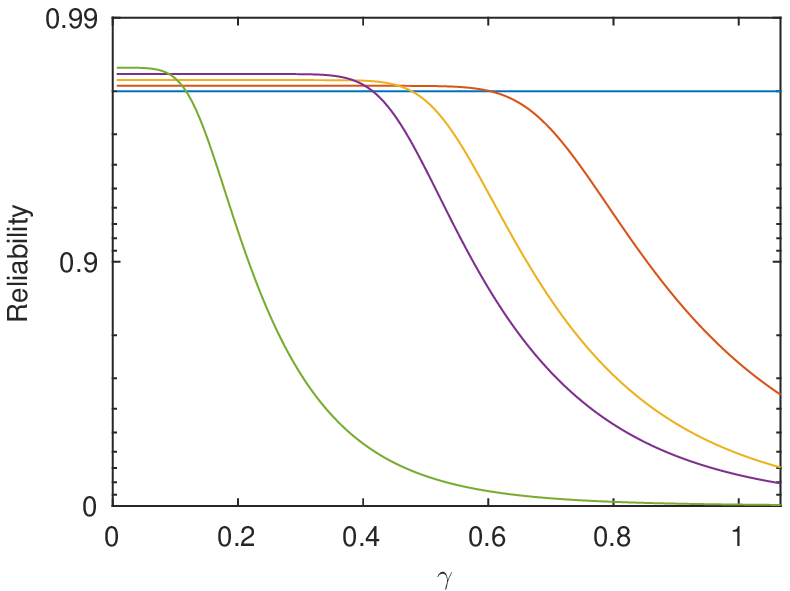}}
    ~ 
    \subfigure[Resulting reliability]{\includegraphics[width=0.4\textwidth]{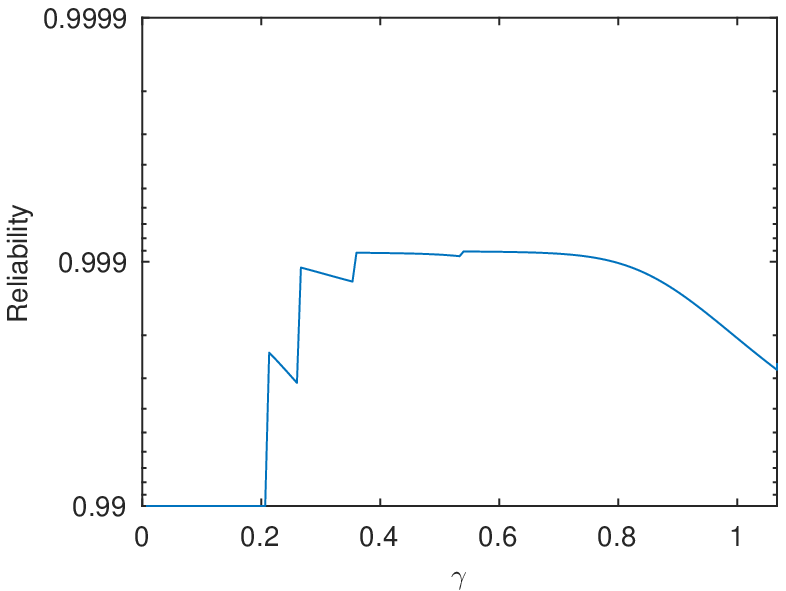}}
    \caption{Example showing the effect of increasing $\gamma$ equally on all interfaces. The example is created for scenario $\mathcal{C}$ in Table \ref{tab:scenarios}, for a latency requirement of $l=0.4$~s.}
    \label{fig:gamma_var_example}
\end{figure}

Assuming that the optimization is solved using a brute-force search, the search space grows as $\left(\sfrac{1}{\delta_\gamma}\right)^N$, where $\delta_\gamma$ is the step size between $\gamma$-values. In practice, the computational tractability of a brute-force search is therefore limited by the number of interfaces $N$ and choice of step size $\delta_\gamma$.
The problem in eq. \eqref{eq:opt_weighted} does not immediately have an analytical solution, since the payload assignment weights in $\bm{\gamma}$ do not translate linearly into specific reliability values. Specifically, when increasing the $\gamma$ value for an interface and thereby increasing the amount of coded payload, the reliability for a specific latency is going to decrease at some point due to the increasing packet size, as exemplified in Fig. \ref{fig:gamma_var_example}(a). 
On the other hand, a combination of two or more interfaces' $\gamma$-values can add up to $\gamma_\text{d}$ and thereby improve the overall reliability (shown in Fig. \ref{fig:gamma_var_example}(b) as jumps in curve), even if the reliability of the individual interfaces decreases as $\gamma$ goes up. 
This behavior, that the overall reliability decreases before it suddenly jumps up, combined with the fact that the $\gamma$ value should be adjusted for each interface individually, narrows the possibilities for analytic solutions.
Therefore, for the numerical results, we include results from a brute-force search that tries out all combinations of $\gamma$-values on the different interfaces, with a step size that is sufficiently coarse to make the search computationally tractable.

\section{Transmissions over interfaces with failure correlation}\label{sec:miftx_correlated}
In systems where the used interfaces cannot be considered completely independent due to common error causes, it is necessary to jointly consider the correlated failure states of the system and the interfaces' latency-reliability functions.
Interfaces with correlated failures could be two Wi-Fi interfaces operating on the same frequency band, thus being affected by the same interference sources, or it could be two cellular interfaces connected to the same base station tower, where, e.g. a power outage would affect both connections.

Correlated failures can be modeled using a \ac{CTMC} state model. For such a model, we calculate the combined latency-reliability function as:
\begin{equation}
		F_{m\text{-dep}}(x,B) = \sum\limits_{s=1}^{L} \pi_s \cdot H_s(x,B), \label{eq:F_dep}
\end{equation}
where $L$ is the number of states in the \ac{CTMC}, $\pi_s$ is the steady-state probability of state $s$ in the \ac{CTMC}, and $H_s(x,B)$ characterizes the latency-reliability function of state $s$.

Since the latency-reliability function $H_s(x,B)$ associated with a given system state $s$ depends on the actual transmission strategy, we will in following consider a specific case study from which it becomes clear how $H_s(x,B)$ is computed.

\subsection{Case study: Correlated failures in three-interface system}
\begin{figure}[htb]
	\centering
	\includegraphics[width=\linewidth]{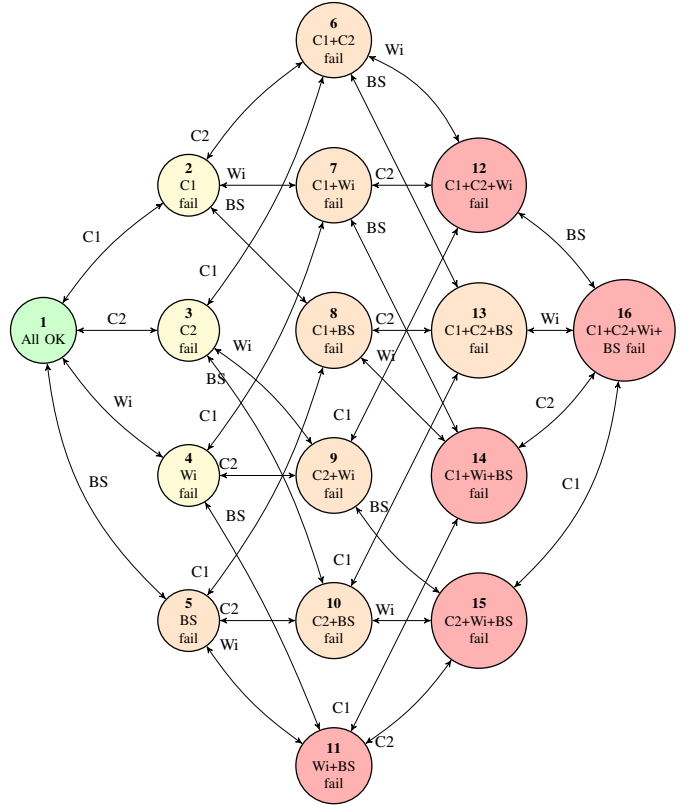}
	\caption{CTMC model of states in the three interface system. Colors indicate the number of interfaces \textit{up/down} as: Green: 3/0, yellow: 2/1, orange: 1/2, red: 0/3. An arrow represents a failure rate in the right direction and restoration rate in the left direction, e.g., $\lambda_{C1}$ and $\mu_{C1}$ between states 1 and 2.}
	\label{fig:15-state_diagram}
\end{figure}

We assume that an M2M device in Fig. \ref{fig:system_model} (b) is connected with Wi-Fi to a fiber connection and by two cellular interfaces, denoted by C1 and C2. This is an example of a mission critical MTC use case from smart grid systems \cite{stefanovic2014sunseed}.
If the cellular interfaces C1 and C2 belong to the same operator, then they are likely located at the same base station tower, such that we need to take into account the probability of the cellular links failing simultaneously due to common error causes.
The CTMC in Fig. \ref{fig:15-state_diagram} shows the different modes of operation considered for the case study. In addition to the independent failures of C1, C2, and W-Fi the model also includes BS failures. In states with BS failure, namely states 5, 8, 10, 11, 13, 14, 15, and 16, neither of the two cellular interfaces will be functioning, since the BS failure represents the common error causes that affect both cellular connections, such as power outage or backhaul connection problems. We need to specifically address the degenerate cases of states 11, 12, 14, 15, and 16 when nothing is functioning.
While the presented failure model is quite simple and only considers the mentioned four high-level failures. This is however sufficient for the needs of this analysis, since the model can be used to determine the most suitable transmission strategy for a certain system configuration and answer what-if questions when having different probabilities of failure correlations.

For each of the considered transmission strategies, we present a short description and define the state-specific latency-reliability functions $H_s(x,B)$ for $s={1,2,\ldots, L}$ that are represented by the vector $\vec{H}(x,B)$.
For compact notation of interface-specific latency-reliability functions, we let $i\!=\!1$ represents $C1$, $i\!=\!2$ is $C2$, and $i\!=\!3$ is Wi-Fi. Further, we define that $\hat{F}_i = F_i(x,\gamma_i B)/A_i$ is the latency-reliability function normalized by the availability $A_i$, thereby making $\hat{F}_i$ a CDF. We do this because we have used $A_i=1-P_\text{e}^{(i)}$ in the parametrization of the \ac{CTMC} model.
By normalizing $P_\text{e}$ out of the latency-reliability function and including it in the CTMC we enable the use of probability theory for the following analysis.
The value of $\gamma_i$ depends on the transmission strategy used, as specified in Table \ref{tab:split_strategies}.
 Illustrations of the strategies and packet size splitting parameters $\gamma_i$, are shown in Table \ref{tab:split_strategies} and they are explained in the next section. Note that the \ac{CTMC} failure model in Fig. \ref{fig:15-state_diagram} is used with all three strategies, since we assume that equipment and transmission failures are independent of the used transmission strategy.

\begin{table}[htb]
\centering
\caption{Packet splitting parameter for different strategies}
\begin{tabular}{lcccccc}
\toprule
\multicolumn{1}{l}{}                 & {\textbf{cloning}}             && {\textbf{2-of-3}}             &                         & {\textbf{weighted}}  \\ \cmidrule{2-2} \cmidrule{4-4} \cmidrule{6-6}
$\gamma_{1}$ & $1$ & & $\sfrac{1}{2}\cdot\gamma_\text{d}$ & & variable \\
$\gamma_{2}$ & $1$ & & $\sfrac{1}{2}\cdot\gamma_\text{d}$ & & $1\!-\!\gamma_1$ \\
$\gamma_{3}$ & $1$ & & $\sfrac{1}{2}\cdot\gamma_\text{d}$ & & $1$ \\
\bottomrule
\end{tabular}
\label{tab:split_strategies}
\end{table}

\subsection{Packet cloning on three interfaces}
For each state in Fig. \ref{fig:15-state_diagram}, we need to specify how the interfaces' latency-reliability functions $F_i(x,B)$ should be combined. In states where more than one interface is available, the latency is given by the first arriving packet when using cloning. Let the independent \acp{RV} $X_1, ..., X_k$ represent the latency of each of the $m\in\{1,2,3\}$ interfaces. The latency CDF of the first arriving is known to be $F_\text{min}=1-\Pi_{j=1}^m(1-F_{j})$. Thus, the $F_i(x,B)$ functions are combined as shown in Table \ref{tab:P_vectors} in order to obtain $H_s(x,B)$ for all $16$ states.
The resulting latency-reliability function is computed using \eqref{eq:F_dep}. 

\begin{table*}[t]
\footnotesize
\centering
\caption{Latency-reliability function vector $\vec{F}^\mathrm{st}(x,B)$ for the considered strategies, with $\hat{F}_i = {F_i(x,\gamma_i B)}/{A_i}$.}
\begin{tabular}{rccccc}
\toprule
&\textbf{cloning} & & \textbf{2-of-3} & & \textbf{weighted (across C1 and C2)} \\ \cmidrule{2-2} \cmidrule{4-4} \cmidrule{6-6}
$\begin{matrix} 
\textit{1} \\ \textit{2} \\ \textit{3} \\ \textit{4} \\ \textit{5} \\ \textit{6} \\ \textit{7} \\ \textit{8} \\ \textit{9} \\ \textit{10} \\ \textit{11} \\ \textit{12} \\ \textit{13} \\ \textit{14} \\ \textit{15} \\ \textit{16}
\end{matrix}$ &
$\begin{bmatrix}
     1-(1-\hat{F}_{1})(1-\hat{F}_{2})(1-\hat{F}_{3}) \\
     1-(1-\hat{F}_{1})(1-\hat{F}_{3}) \\
     1-(1-\hat{F}_{1})(1-\hat{F}_{3}) \\
     1-(1-\hat{F}_{1})(1-\hat{F}_{2}) \\
     \hat{F}_{3} \\
     \hat{F}_{3} \\
     \hat{F}_{2} \\
     \hat{F}_{3} \\
     \hat{F}_{1} \\
     \hat{F}_{3} \\
     0 \\
     0 \\
	 \hat{F}_{3} \\
     0 \\
     0 \\
     0
    \end{bmatrix}$
& &
$\begin{bmatrix}
	 F_\text{A} + F_\text{B} + F_\text{C} + F_\text{D}\\
     \hat{F}_{2}\hat{F}_{3} \\
     \hat{F}_{1}\hat{F}_{3} \\
     \hat{F}_{1}\hat{F}_{2} \\
     0 \\
     0 \\
     0 \\
     0 \\
     0 \\
     0 \\
     0 \\
     0 \\
     0 \\
     0 \\
     0 \\
     0
    \end{bmatrix}$
& &
$\begin{bmatrix}
      1\!-\!(1-(\hat{F}_{1}\hat{F}_{2}))(1-\hat{F}_{3}) \\
     \hat{F}_{3} \\
     \hat{F}_{3} \\
     \hat{F}_{1}\hat{F}_{2} \\
     \hat{F}_{3} \\
     \hat{F}_{3} \\
     0\\
     \hat{F}_{3} \\
     0\\
     \hat{F}_{3} \\
     0\\
     0\\
     \hat{F}_{3} \\
     0\\
     0\\
     0
    \end{bmatrix}$ \\
    \bottomrule
\end{tabular}
\label{tab:P_vectors}
\end{table*}
\renewcommand\arraystretch{1}

\subsection{2-of-3 packet splitting on three interfaces}
As explained in sec. \ref{sec:strategies}, a 2-out-of-3 strategy requires only coded packets corresponding to $\sfrac{1}{2}\cdot\gamma_\text{d}$ of the source packet to be sent on each interface. Consequently, the state-specific latency-reliability functions are different from the ones in packet cloning. 
In state 1, to compute the probability of receiving at least 2 fragments within a latency value $x$, we need to consider all ways this can happen. Either all three fragments are received before $x$ or any two of the three fragments are received before $x$.
The CDFs of these four cases, arbitrarily named A--D, are:
\begin{equation}
	\begin{matrix*}[l]
		F_\text{A} = \hat{F}_1\hat{F}_2\hat{F}_3 &\quad&  F_\text{B} = \hat{F}_1\hat{F}_2(1-\hat{F}_3)\\
		F_\text{C} = \hat{F}_1(1-\hat{F}_2)\hat{F}_3 &\quad& F_\text{D} = (1-\hat{F}_1)\hat{F}_2\hat{F}_3
	\end{matrix*}
\end{equation}
For the CDF of state 1 we use their sum as shown in Table \ref{tab:P_vectors}. On a side note, notice that if we have identical interfaces such that $F_x=\hat{F}_1=\hat{F}_2=\hat{F}_3$ then the expression for the CDF of state 1 simplifies to:
$$3 F_x^2 (1-F_x) + F_x^3,$$
which equals the formula for reliability of a 2-out-of-3 system \cite{rausand2004system}.
For states $2-4$, we use that the second fragment is the last and that its latency CDF is $F_\text{max}=\Pi_{j=1}^m(F_j)$ \cite{ross1996stochastic}.


\subsection{Weighted packet splitting on three interfaces}
In this case study, we consider a particular configuration, where we send a full copy of the message via Wi-Fi and split another copy between the two cellular interfaces, to achieve a latency reduction.
As this situation is identical to the situation analyzed in sec. \ref{sec:analysis}, we can directly use the derived expression in \eqref{eq:analytic_splitting} to give the optimal $\gamma$-values.
 
The latency-reliability function vector $\vec{H}(x,B)$ for the \textit{weighted} strategy is shown in Table \ref{tab:P_vectors} and here we use that the CDF of the latency of the last arriving is $F_\text{max}=\Pi_{j=1}^m(F_{j})$.

\section{Numerical results}
\label{sec:results} 

For the numerical results we will consider the different scenarios specified in Table \ref{tab:scenarios}. For each scenario, one or more latency requirements are specified. These latency requirements have been selected so as to demonstrate the potential gains of optimization.
The considered technologies are using the reliability specifications shown in Table \ref{tab:pl_to_rss_and_reliability}.

%

\setlength\tabcolsep{5pt}
\begin{table}[bt]
	\centering
	\caption{Linear regression parameters from RTT measurements and assumed reliability values of equipment.}
\label{tab:pl_to_rss_and_reliability}
	\begin{tabular}{lccccccc}
	\toprule
		 			& GPRS & EDGE & UMTS & HSDPA & LTE & Wi-Fi & BS \\ \cmidrule{2-8}
	 	$\alpha$ 	& 0.70 & 0.46 & 0.43 & 0.35 & 0.0067 & 0.00068 & - \\
		$\beta$  	& 400 & 230 & 200 & 178 & 41 & 2.3 & - \\
		$P_\text{e}$& 0.984 & 0.983 & 0.982 & 0.981 & 0.980 & 0.950 & 0.990 \\ \bottomrule
	\end{tabular}
\end{table}
\setlength\tabcolsep{6pt}

\begin{table*}[bt]
	\centering
    \footnotesize
	\caption{Interface and parameter specifications of scenarios $\mathcal{A}$, $\mathcal{B}$, $\mathcal{C}$, and $\mathcal{D}$.}
	\label{tab:scenarios}
	\begin{tabular}{ccccccccccc}
	\toprule
			& IF1 	& IF2 	& IF3 	& IF4 	& IF5 	& & $B$ & & $\bm{l}$ 	& $\bm{w}$ 	\\ \cmidrule{2-6} \cmidrule{8-8} \cmidrule{10-11}
$\mathcal{A}$  	& Wi-Fi	& UMTS & GPRS 	& - 	& - 	& & 1500 bytes & & $[0 \ldots 1]$ s		& $[0 \ldots 1]$\\ 
$\mathcal{B}$ 	& Wi-Fi	& UMTS & EDGE 	& GPRS 	& - 	& & 3500 bytes & & $[0.7]$ s		& $[1]$		\\ 
$\mathcal{C}$	& LTE 	& HSDPA & UMTS 	& EDGE 	& GPRS 	& & 1500 bytes & & $[0.1, 0.4, 0.9^*]$ s 	& $[1, 10, 100^*]$	\\
$\mathcal{D}$	& HSDPA	& HSDPA	& GPRS 	& GPRS 	& GPRS 	& & 1500 bytes & & $[0.5]$ s 		& $[1]$		\\
$\mathcal{E}$	& Wi-Fi	& UMTS	& EDGE 	& - 	& - 	& & 1500 bytes & & $[0 \ldots 1]$ s		& $[0 \ldots 1]$\\  \bottomrule
	\end{tabular}
\end{table*}

\begin{figure}[htb]
	\centering
	\includegraphics[width=\figscl\linewidth]{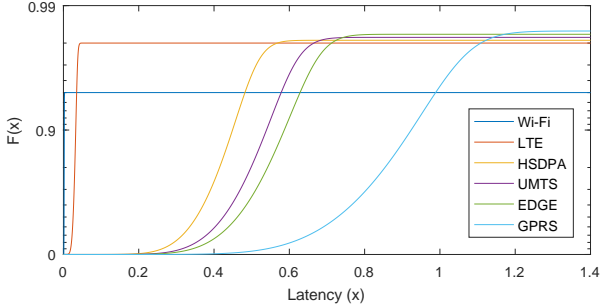}
	\caption{Latency-reliability curves $F_i(x,B)$ for all considered technologies for $B=1500$~bytes.}
	\label{fig:cdfs_ifs}
\end{figure}

While the distribution of latency measurements is usually long-tailed \cite{borella1997self,jacko2000effect}, we will for simplicity use the normal probability distribution to generate latency distributions in the numerical results. Notice that while the used probability distribution of course influences the specific results, the methods and general tendencies presented in this paper does not change.
 Specifically, we assume that the latency of transmissions of packet size $B$ through a specific interface/path Gaussian distributed with mean $\mu$ defined as:
\begin{equation}\label{eq:latency_B}
	\mu = \frac{\alpha \cdot B + \beta}{2} [ms]
\end{equation}
and due to lack of information about the distribution, we assume $\sigma = \frac{\mu}{10}~[ms]$.
The parameters $\alpha$ and $\beta$ characterize the assumed linear relationship between packet size and delay for an interface. The values of $\alpha$ and $\beta$ are shown in Table \ref{tab:pl_to_rss_and_reliability}. The values of $\alpha$ and $\beta$ are derived from field measurements conducted by Telekom Slovenije within the SUNSEED project \cite{sunseed2014web}, whereas the $P_\text{e}$ values are arbitrarily chosen.
The resulting latency-reliability characteristics are shown in Fig. \ref{fig:cdfs_ifs} for the case of $B=1500$~bytes\footnote{Note that with smaller values of $B$, the curves shift towards the left.}.

\subsection{Independent interfaces}
\begin{figure}[htb]
	\centering
	\includegraphics[width=\figscl\linewidth]{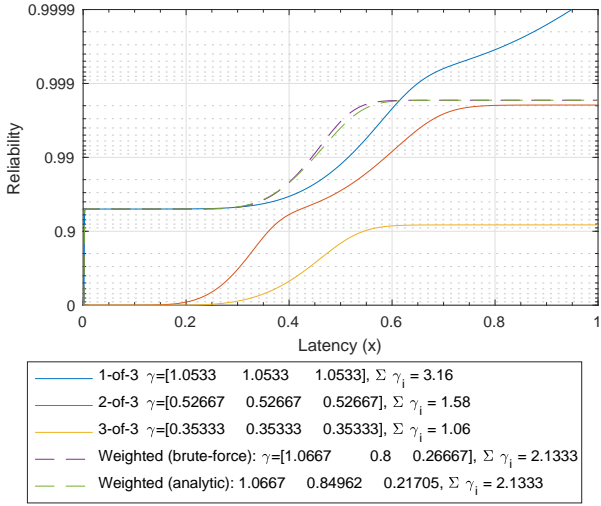}
	\caption{Reliability results for scenario $\mathcal{A}$.}
	\label{fig:scenarioA}
\end{figure}

Initially, we study the simple scenario $\mathcal{A}$, for which we solved the weighted splitting between two interfaces analytically in sec. \ref{sec:analysis}. That is, we used \eqref{eq:analytic_splitting} to determine the optimal splitting threshold $\gamma$. Notice that $\bm{l}$ and $\bm{w}$ are parametrized so that the numerical optimization calculates the expected latency like the analytical optimization.
The results are shown in Fig. \ref{fig:scenarioA}, and show a visually good correspondence between the analytical result and the brute-force search. The brute-force search has a slightly lower expected latency, due to the weight assignment being different. We attribute this minor difference to the use of the approximation of $\mathbb{E}[ \max (X_{A},X_{B}) ]$ from \cite{clark1961greatest}.

In relation to the general idea of splitting, the most important question we seek to answer, is if it makes sense to spend the additional effort required to find the optimal $\gamma$-values for a weighted splitting or if it suffices to use one of the simpler $k$-out-of-$N$ strategies. It is intuitively clear that if the used technologies are all identical, then a $k$-out-of-$N$ strategy will be optimal. But how much better is a weighted scheme in a heterogeneous scenario? To answer this we study three different scenarios that are specified in Table \ref{tab:scenarios}.

\begin{figure}[htb]
	\centering
	\includegraphics[width=\linewidth]{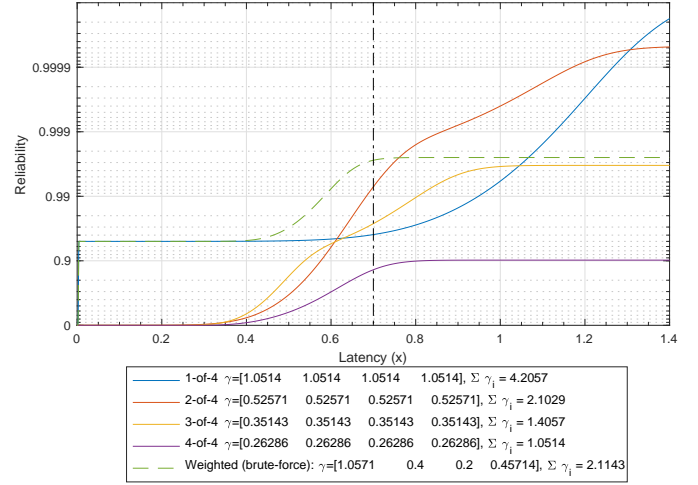}
	\caption{Reliability results for scenario $\mathcal{B}$.}
	\label{fig:scenarioB}
\end{figure}

The resulting reliabilities for the different transmission strategies are shown for scenario $\mathcal{B}$ in Fig. \ref{fig:scenarioB}. The most distinctive observation is that in the low latency region $x<0.3$~s, only the 1-out-of-4 and Weighted strategies provide any reliability. However, around the target latency $x=0.7$~s, both the 2-out-of-4 and 1-out-of-4 strategies achieve higher reliability than the 1-out-of-4 since the payload is split between the interfaces. Nevertheless, the optimal weight assignment used by the Weighted strategy has the highest reliability at $x=0.7$~s. The assigned $\gamma$-values are shown in the figure legend. 
In comparison to the 1-out-of-4 (Cloning) strategy we see a significant improvement in reliability from 0.95 to 0.997 at the target latency $x=0.7$~s. In terms of latency, at R=0.997, we see a reduction from 1.05~s to 0.7~s.

\begin{figure}[htb]
	\centering
	\includegraphics[width=\linewidth]{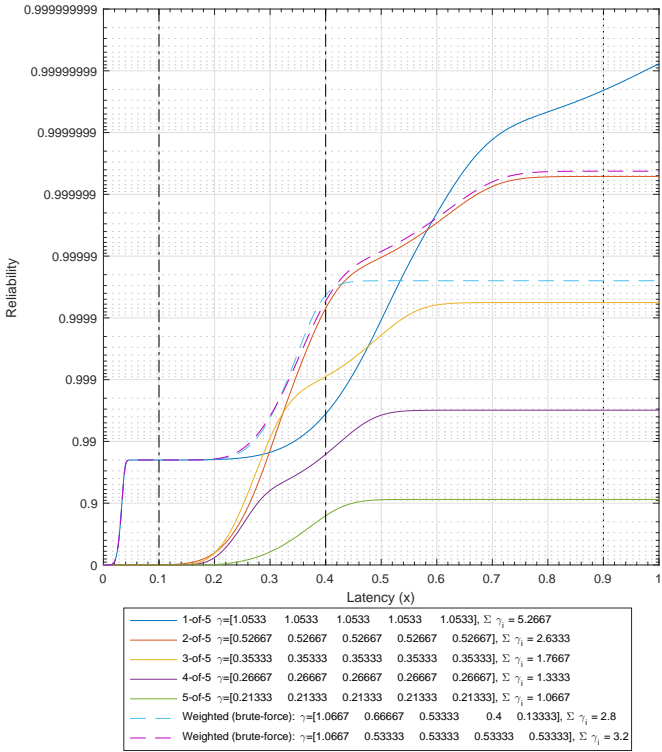}
	\caption{Reliability results for scenario $\mathcal{C}$. Note: the target latency $l_2=0.9$~s only applies to the last strategy.}
	\label{fig:scenarioC}
\end{figure}

While scenario $\mathcal{B}$ demonstrated how latency can be lowered, the results for scenario $\mathcal{C}$ in Fig. \ref{fig:scenarioC} show two examples of latency-reliability trade-offs that are achieved by considering both when the starred $l$ and $w$ values in Table \ref{tab:scenarios} are included and excluded. In both cases the weighted strategy achieves some reliability in the low latency region ($x<0.2$~s) similar to the 1-out-of-5 strategy and it has the reliability of the 2-out-of-5 strategy around $x=0.4$~s. The difference between the 2 results is that the last one transmits more redundancy data and achieves higher reliability in the $x>0.4$~s region.

\begin{figure}[htb]
	\centering
	\includegraphics[width=\linewidth]{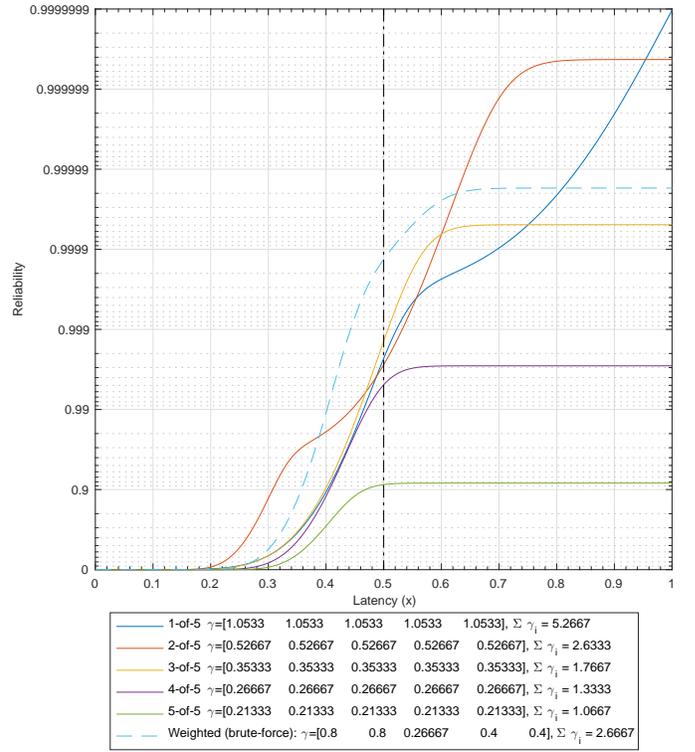}
	\caption{Reliability results for scenario $\mathcal{D}$.}
	\label{fig:scenarioD}
\end{figure}

The last results concerning scenario $\mathcal{D}$ that are shown in Fig. \ref{fig:scenarioD} are interesting since they demonstrate a more mixed data allocation. This results in the reliability at $x=0.5$~s being 0.9999, which is one decade better than any $k$-out-of-$N$ strategy that only go up to 0.999.

\begin{table}[bt]
	\centering
	\caption{Case study failure and restoration rates}
	\begin{tabular}{lcc}
	\toprule
		 				& $\lambda$ (f/week) & $\mu$ (r/week) \\ \cmidrule{2-3}
		Wi-Fi + Fiber (Wi) 	& 1.47	 	 & 28 (6 hrs/r) \\
		Cellular (C1, C2)	& 0.64		 & 50.4 (200 min/r) \\
		Base station (BS) 	& 0.76 	 	 & 50.4 (200 min/r) \\ \bottomrule
	\end{tabular}
	\label{tab:f_and_r_rates}
\end{table}

\subsection{Interfaces with failure correlation}
For this case study, we consider that besides failing independently, C1 and C2 can also fail simultaneously due to a common BS failure. This will be reflected in the \emph{MC model} results, whereas the \emph{independent} results do not account for common cause failures.

For evaluating the resulting performance of the considered transmission modes, actual data on \ac{MTTR} and availability levels of different technologies has been used. From these numbers, the unspecified failure and restoration rates have been determined. The approach to parametrize the CTMC model is explained in the Appendix.
Table \ref{tab:f_and_r_rates} presents the used failure and restoration rates.

\begin{figure}[htb]
	\centering
	\includegraphics[width=\figscl\linewidth]{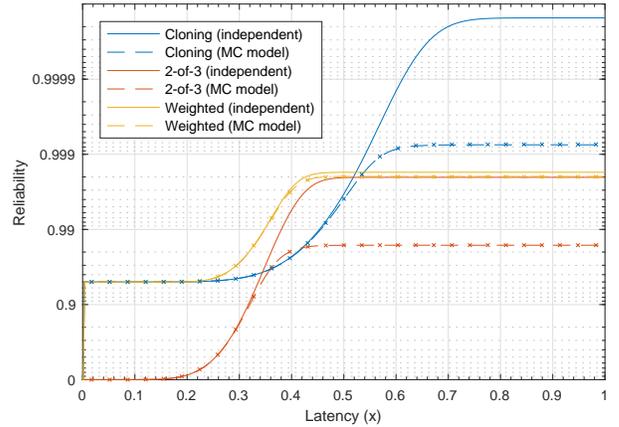}
	\caption{Reliability results for scenario $\mathcal{E}$, where we found $\gamma=0.55$.}
	\label{fig:scenarioE}
\end{figure}

With failure and restoration rates fully specified, the resulting latency-reliability performance is calculated using the methods outlined in sec.~\ref{sec:reliability_miftx}. The different model results have been verified using Matlab-based simulation. We first simulated the transitions between states in the \ac{CTMC} model in Fig. \ref{fig:15-state_diagram} with exponential sojourn times given from the rates in Table \ref{tab:f_and_r_rates}.
Hereafter we replayed the state sequence and for every 1 min simulation time, a random Gaussian latency value was drawn for the interfaces available in the current state.  Depending on the required packet fragments of the strategy either a transmission latency or timeout value resulted. The CDF of these values is shown with crosses in Fig. \ref{fig:scenarioE}.

In all plots in Fig. \ref{fig:scenarioE} we see that the \textit{Cloning} strategy, which uses three times as much bandwidth as a single-interface transmission, achieves the highest reliability in the high latency region. The impact of failure correlations is shown from the difference between the \emph{independent} and \emph{MC model} curves. For cloning, the difference amounts to more than one decade at high latency values. This difference results from the fact that both cellular interfaces are depending on the base station being operational. That is, in cases where the base station fails (model states: 5, 8, 10, 11, 13, 14, 15, 16) neither C1 or C2 will be operational.
\footnote{The used value of $P_\text{e}=0.99$ for the base station may be high compared to a real-life system, however the main point of the analysis is to show how such factors can be modeled.}
The 2-out-of-3 strategy uses only half the total bandwidth of cloning. However, the dependence on at least two working interfaces causes the lack of reliability before $0.2$~s. While in the independent case the 2-out-of-3 strategy is able to reduce latency up to almost 0.9~s at $R=0.996$, the MC model result that accounts for correlated failures, is the worst strategy, except the small interval $0.35-0.42$~s where it reaches the same reliability level as cloning.
Finally, the \textit{Weighted} strategy shows the best performance for low latency ($x \leq 0.5$~s), whereas it is only slightly worse than \emph{Cloning (independent)} for higher latency values. It is worth noticing that the difference between the \emph{independent} and  \emph{MC model} results for this strategy is minimal. We explain this from the fact that in the weighted strategy the cellular interfaces are inherently depending on each other also in the independent case, whereas for cloning and 2-out-of-3, the two cellular interfaces are independent in the independent case, but dependent when using the MC model.

\begin{figure}[htb]
	\centering
	\includegraphics[width=\figscl\linewidth]{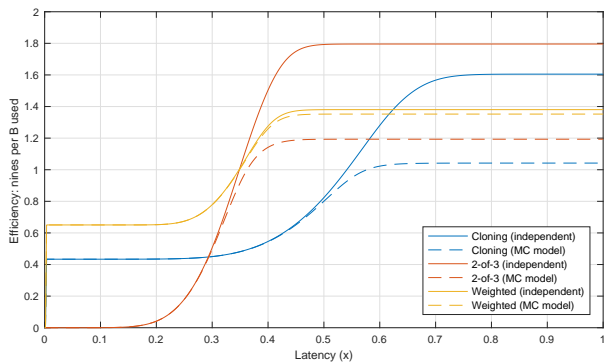}
	\caption{Efficiency results for scenario $\mathcal{E}$.}
	\label{fig:scenarioE_eff}
\end{figure}

Besides considering only the level of reliability that each strategy can achieve, we are showing also the efficiency as the achieved reliability (in number of nines) in relation to the amount of coded data transmitted ($B$ bytes) in Fig. \ref{fig:scenarioE_eff}. While the independent results show that both the 2-out-of-3 and cloning strategies are better than the weighted strategy, this observation does not hold for the case with correlated failures (MC model). In this case the weighted strategy is the best choice for the whole span of latency values.

\subsection{Experimental validation}\label{sec:exp_results}
\begin{figure}
    \centering
    \subfigure[]{\includegraphics[width=0.4\textwidth]{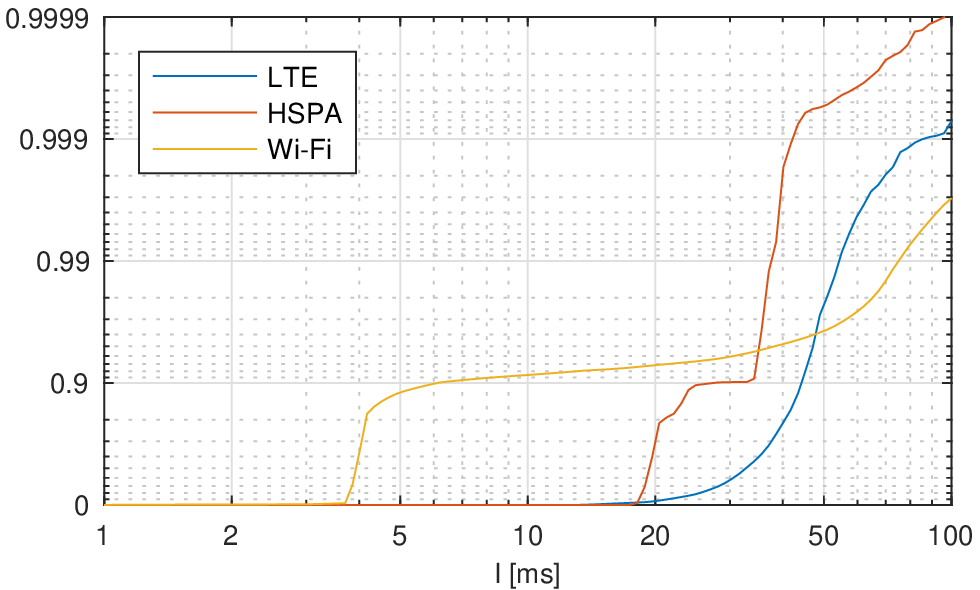}}
    ~ 
    \subfigure[]{\includegraphics[width=0.4\textwidth]{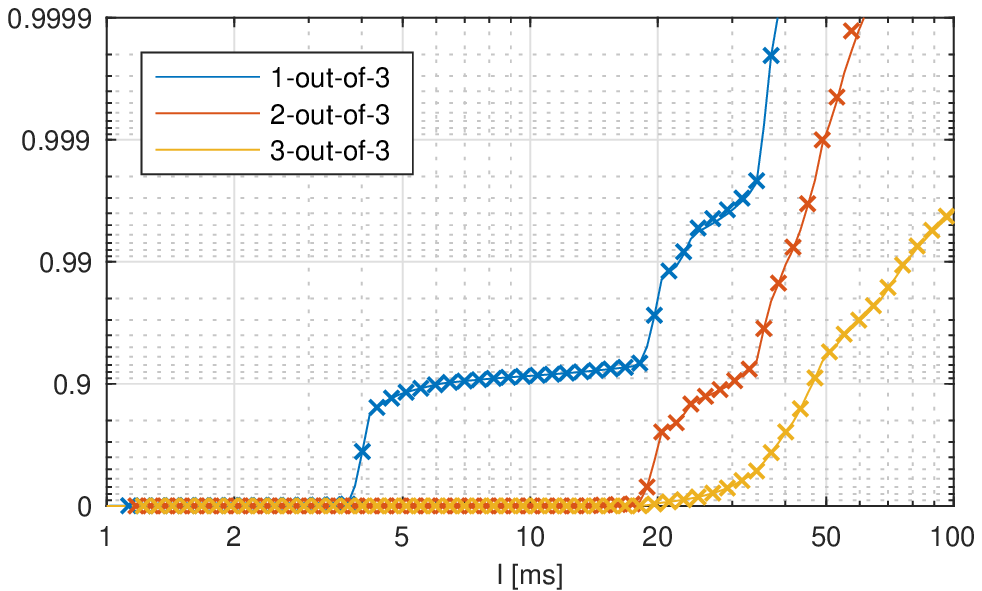}}
    \caption{(a) Interfaces' latency-reliability curves. Wi-Fi is IEEE 802.11n. (b) Resulting performance of considered strategies. Lines show the results computed using the method presented in sec. \ref{sec:reliability_miftx}, crosses show the results of playback-simulation.}
    \label{fig:experimental}
\end{figure}

In addition to the theoretical and model-based results presented above, we have also validated the proposed method for combining the latency-reliability functions using experimental results. While this validation experiment does not explicitly cover all of the scenarios $\mathcal{A}$-$\mathcal{E}$, it shows how well the latency-reliability curves can be used to calculate the actual performance of multi-interface transmissions.
In the experiment, we have used traces of latency measurements for different communication technologies. Such traces were obtained by sending small (128~bytes) UDP packets every 100~ms between a pair of GPS time-synchronized devices through the considered interface (LTE, HSPA, or Wi-Fi) during the course of a work day at Aalborg University campus. Each trace file can thus be used to play back a time sequence of one-way end-to-end latencies. Our experimental results of multi-interface transmissions are obtained by playing back the three trace files at the same time time in a simulation, where for every 100 ms, the outcome of each considered strategy is recorded. When the simulation is done, a latency-reliability curve is calculated for each strategy as the cdf of the recorded outcomes in each 100 ms timestep. This is shown with crosses in Fig. \ref{fig:experimental} (b). The validation consists in comparing these results to the results that are obtained by using the curves in Fig. \ref{fig:experimental} (a) to compute the resulting latency-reliability curves using the methods described in sec. \ref{sec:reliability_miftx}. Those results are shown as lines in Fig. \ref{fig:experimental} (b).

From the results in Fig. \ref{fig:experimental}, we see how the 1-out-of-3 strategy is able to outperform any individual interface, as expected.
The plot does not include any result for the Weighted scheme, since the small payload size does not allow for any gain through payload splitting. The lines that represent the theoretical calculation of performance are practically coinciding with the crosses representing the experimental results. This shows that the methods for calculating the resulting performance by relying on the latency-reliability curves of the interfaces, as described in Sec. \ref{sec:reliability_miftx}, indeed produces accurate results when used with actual traffic traces.

\section{Conclusions and Outlook}
\label{sec:conclusion} 
It is expected that 5G will integrate various communication technologies to support ultra-reliable and low latency (URLLC) use cases. In this work we denote this integration \emph{interface diversity} and consider different strategies for utilizing multiple interfaces simultaneously, to achieve high reliability and low latency. By flexibly allocating coded fragments of the encoded payload message to different interfaces, according to their bit-rate, latency and reliability properties, it becomes possible to trade-off transmission latency and reliability. We have considered both static $k$-out-of-$n$ strategies and optimized weighted strategies.

For evaluating performance, we have proposed an analysis framework that combines traditional reliability models with technology-specific latency probability distributions. The proposed models can be used both for systems with independently failing communication paths and for systems with common error causes, e.g. if cellular technologies reside in the same base station tower.

Our main findings are that 1) interface diversity strategies can lower the latency up to around $40\%$ in practical systems when large messages are transmitted using low rate technologies, where the time to transmit the bits over the air is substantial in relation to the access delay; 2) in some cases only the optimized weighted strategy (and not the simple $k$-out-of-$n$) can deliver latency reduction and reliability at low latencies;  3) the optimized weighted strategy enables the fine-tuning of the latency-reliability trade-off for a specific scenario; and 4) we have experimentally validated the proposed method of computing the resulting performance, and demonstrated the practical gains of interface diversity in a three interface scenario for the $k$-out-of-$n$ strategies.

\section*{Acknowledgment}
This work is partially funded by EU, under Grant agreement no. 619437. The SUNSEED
project is a joint undertaking of 9 partner institutions and their contributions are
fully acknowledged. The work was also supported in part by the European Research Council (ERC Consolidator Grant no. 648382 WILLOW) within the Horizon 2020 Program.

Also, thanks to Kasper~F.~Trillingsgaard for constructive comments and suggestions.

\begin{figure}[htb]
	\centering
	\includegraphics[width=0.8\linewidth]{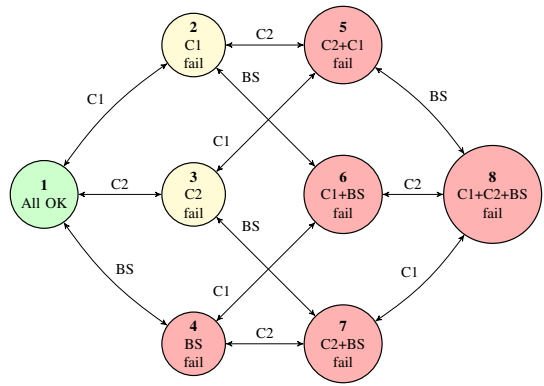}
	\caption{State-transition diagram of the continuous time Markov chain that represents the cellular connections C1 and C2 with correlated failures.}
	\label{fig:5-state_diagram}
\end{figure}

\appendix
This appendix explains the approach used to determine the Markov chain failure and restoration rates for the dependent cellular technologies C1 and C2. For this, we consider the CTMC model corresponding to the cellular subsystem of Fig. \ref{fig:system_model} (b). This subsystem is shown in Fig. \ref{fig:5-state_diagram}.

Initially, we specify the known individual availabilities $A_\text{C1}$ and $A_\text{C2}$ as well as the known base station availability $A_\text{BS}$, given in Table \ref{tab:f_and_r_rates}. 
Transitions between states are specified by the failure rates denoted by $\lambda$ and restoration rates denoted by $\mu$. Notice that neither failure rates or restoration rates are known for the considered case study. We have therefore made assumptions in the values of the restoration rates as specified in Table \ref{tab:f_and_r_rates}. 

Given the availabilities $A_\text{C1}$, $A_\text{C2}$, and $A_\text{BS}$, we determine the state probabilities $\pi_i$ of the states in Fig. \ref{fig:5-state_diagram}, by solving the following linear equation system that explains the relations between the steady state probabilities and availability probabilities:
%
\begin{equation}
 	\begin{bmatrix*}[r]
	1 & 0 & 1 & 1 & 0 & 0 & 1 & 0 \\
	1 & 1 & 0 & 1 & 0 & 1 & 0 & 0 \\
	1 & 1 & 1 & 0 & 1 & 0 & 0 & 0 \\
	0 & 1 &-1 & 0 & 0 & 0 & 0 & 0 \\
	0 & 0 & 0 & 0 & 1 &-1 & 0 & 0 \\
	0 & 0 & 0 & 0 & 0 & 1 & 0 & 0 \\
	0 & 0 & 0 & 0 & 0 & 0 & 1 & 0 \\
	0 & 0 & 0 & 0 & 0 & 0 & 0 & 1 \\
	1 & 1 & 1 & 1 & 1 & 1 & 1 & 1
    \end{bmatrix*}
	\begin{bmatrix}
	\pi_1\\
	\pi_2\\
	\pi_3\\
	\pi_4\\
	\pi_5\\
	\pi_6\\
	\pi_7\\
	\pi_8
    \end{bmatrix}
 =	\begin{bmatrix}
	A_\text{1}\\
	A_\text{2}\\
	A_\text{BS}\\
	0\\
	0\\
	\bar{A}_\text{C1}\bar{A}_\text{BS}\\
	\bar{A}_\text{C2}\bar{A}_\text{BS}\\
	\bar{A}_\text{C1}\bar{A}_\text{C2}\bar{A}_\text{BS}\\
	1
    \end{bmatrix},\nonumber
\end{equation}
%
where $\bar{A}_*=1-A_*$ is used for compact notation.

Having obtained the state probabilities $\boldsymbol{\pi}=[\pi_1 \ldots \pi_5]$, we set up the following balance equations that explain the relations between the failure and restoration rates according to Fig. \ref{fig:5-state_diagram}. The assumed mean restoration rates in Table \ref{tab:f_and_r_rates} are given as input and we can then solve the corresponding linear system:
%
\begin{equation}
 	\begin{bmatrix*}[r]
    -\pi_1 & -\pi_1 & -\pi_1 & \pi_2 & \pi_3 & \pi_4 \\
    \pi_1 & -\pi_2 & -\pi_2 & -\pi_2 & \pi_4 & \pi_5 \\
    -\pi_3 & \pi_1 & -\pi_3 & \pi_4 & -\pi_3 & \pi_5 \\
    -\pi_4 & -\pi_4 & \pi_1 & \pi_6 & \pi_7 & -\pi_4 \\
    \pi_3 & \pi_2 & -\pi_5 & -\pi_5 & -\pi_5 & \pi_8 \\
    \pi_4 & -\pi_6 & \pi_2 & -\pi_6 & \pi_8 & -\pi_6 \\
    -\pi_7 & \pi_4 & \pi_3 & \pi_8 & -\pi_7 & -\pi_7 \\
    \pi_7 & \pi_6 & \pi_5 & -\pi_8 & -\pi_8 & -\pi_8 \\
    -1 & 1 & 0 & 0 & 0 & 0 \\
    0 & 0 & 0 & -1 & 1 & 0 \\
    0 & 0 & 0 & 1 & 0 & 0 \\
    0 & 0 & 0 & 0 & 1 & 0 \\
    0 & 0 & 0 & 0 & 0 & 1 
    \end{bmatrix*}
	\begin{bmatrix}
	\lambda_\text{C1}\\
	\lambda_\text{C2}\\
	\lambda_\text{BS}\\
	\mu_\text{C1}\\
	\mu_\text{C2}\\
	\mu_\text{BS}\\
    \end{bmatrix}
 =	\begin{bmatrix}
	0 \\
	0 \\
	0 \\
	0 \\
	0 \\
	0 \\
	0 \\
	0 \\
	0 \\
	0 \\
	\mu_\text{C1} \\
	\mu_\text{C2} \\
	\mu_\text{BS}
    \end{bmatrix}. \nonumber
\end{equation}

Thereby we obtain a set of failure rates $\lambda_\text{C1}$, $\lambda_\text{C2}$, and $\lambda_\text{BS}$ that satisfy the constraints of the system in terms of state probabilities, restoration rates, and balance relations between states.

\bibliographystyle{IEEEtran}

\end{document}